\documentstyle[prl,aps,preprint,epsfig]{revtex} 
\tightenlines
\include{epsf}
\epsfverbosetrue 
\begin{document}
\draft
\title{Efficient Real Space Solution of the Kohn-Sham Equations
with Multiscale Techniques}
\author{
Jian Wang \\
Department of Chemistry \\
University of California \\
Davis, CA 95616, USA\\
.  \\
Thomas L. Beck \\
Department of Chemistry \\
University of Cincinnati \\
Cincinnati, OH 45221-0172, USA \\}

\maketitle

\begin{abstract}

We present a multigrid algorithm for self consistent
solution of the Kohn-Sham
equations in real space.  The entire problem is discretized
on a real space mesh with a high order finite difference
representation. The resulting self consistent equations are
solved on a heirarchy of grids of increasing resolution
with a nonlinear Full Approximation Scheme, 
Full Multigrid
algorithm.  The self consistency is effected by updates of 
the Poisson equation and the exchange correlation potential
at the end of each eigenfunction 
correction cycle. The algorithm leads
to highly efficient solution of the equations, whereby the 
ground state electron distribution is obtained in only 
two or three self consistency iterations on the finest
scale. 

\end{abstract}

\pacs{PACS numbers:31.15.Ar, 71.15.-m, 71.15.Nc, 02.70.-c }

\narrowtext

The plane wave pseudopotential method has proven to be an 
excellent computational strategy for solving large scale
electronic structure problems in condensed 
phases\cite{cp,payne}.  Notable
strengths of the method are the ability to use fast 
Fourier transforms for updating the self consistency 
equations, lack of dependence of the basis on atom positions,
and the clear control of convergence with the cutoff energy
determined by the shortest wavelength mode.  However, the 
method can encounter difficulties in treating widely varying
length scales. The problems are especially severe
for surfaces, clusters, or systems where the pseudopotential
varies rapidly near the nucleus such as first row elements
or transition metals.  Also, charged systems 
present significant complications in a 
plane wave code.  In recent years, 
considerable effort has been
focused on alternative real space
approaches utilizing finite elements or finite difference representations\cite{wwt,chel,davstad,gygi,bernholc,ancilotto,kaxiras,beck}.  
Advantages of these approaches include 
the ability to handle finite or periodic 
systems with equal effort and 
the locality of each
iteration step. Locality leads to simplicity in
developing domain decomposition parallel 
algorithms. In addition, it is relatively straightforward
to implement adaptive grid refinement techniques in order
to focus effort in spacial regions with large variations in the 
computed functions, for example near the nuclei in electronic
structure computations\cite{gygi,tbce}.  
Finally, representation directly
in real space allows for the use of multigrid (MG) algorithms
with their excellent convergence characteristics
and scaling properties\cite{abeval}. 

Several groups have employed MG algorithms for various
portions of the solution process for self 
consistent electronic structure computations.  
White, {\it et al.} \cite{wwt}, developed a finite elements
method for electronic structure which utilized an MG
solver for the Poisson equation.  Davstad\cite{davstad} 
discretized the Hartree-Fock equations for diatomic 
molecules and solved the resulting two dimensional 
equations with a combination of the MG method and
a Krylov subspace method for the coarsest grid
equations.  Gygi and Galli\cite{gygi} proposed an adaptive 
coordinate approach which places increased 
resolution near the nuclei using curved grids. 
They solved the Poisson portion of the problem
with MG techniques.  Briggs, 
{\it et al.}\cite{bernholc} have
developed an MG solver for the Kohn-Sham (KS)
equations which utilizes MG methods for both the
eigenvalue and Poisson problems. Their MG 
eigenvalue solver uses a double discretization
approach and a linearized version of the MG
method. Recently, Ancilotto, 
{\it et al.}\cite{ancilotto}
have presented a similar method for solution 
of the KS equations, and have applied the method
to calculations on charged lithium clusters.
They show that the MG approach is 
as accurate and more efficient
than the corresponding Car-Parrinello method.  
Modine, {\it et al.}\cite{kaxiras} 
developed an adaptive grid
real space method which employs MG preconditioning
for solution of the Poisson equation.  Finally,
we have developed high order MG methods for 
solving the KS equations including all the 
electrons\cite{beck}.  
Typical of the MG solvers to date
is the requirement of 20 or more self consistency 
cycles to obtain convergence. In
this letter, we present a high order multigrid algorithm 
for solving the self consistent Kohn-Sham equations which 
obtains convergence in an order of magnitude less
numerical effort.  The method utilizes the nonlinear 
Full Approximation Scheme (FAS), full multigrid (FMG) 
eigenvalue technique of Brandt, 
{\it et al.}\cite{abeval} with the
further inclusion of new methods for the self consistency
portion of the problem.  

The first step in development of the MG algorithm is 
the discretization of the problem in real space. Here 
we employ a high order finite difference representation. 
This approach has been shown to yield accurate
results in pseudopotential and all electron 
calculations\cite{chel,beck}.
Consider the KS equations\cite{ks} 
in the Local Density 
Approximation (LDA):

\begin{equation}
[- \frac{1}{2} \nabla^2 + v_{eff}({\bf r}) ] \psi_i({\bf r}) = \epsilon_i \psi_i({\bf r}).
\label{eq:ks}
\end{equation}

\noindent
The one-electron effective potential is:
 
\begin{equation}
 v_{eff}({\bf r}) = v_{es}({\bf r})+ v_{xc}(\rho({\bf r})), 
\end{equation}

\noindent
with the electrostatic portion of the potential (nuclear
potential plus Hartree potential) given by:

\begin{equation}
v_{es}({\bf r}) =\sum_{i}\frac{Z_i}{| {\bf r} - {\bf R}_i |} + \int \frac{\rho({\bf r'})}{| {\bf r} - {\bf r'} |} d {\bf r'}.
\end{equation}

\noindent
This potential can be obtained by real space numerical
solution of the Poisson equation:

\begin{equation}
\nabla^2 v_{es}({\bf r}) = - 4 \pi \rho_{tot} ({\bf r}),
\label{eq:poisson}
\end{equation}

\noindent
where $\rho_{tot}$ is the total charge density on the grid 
due to the electrons {\it and} nuclei.  
The exchange-correlation potential in our 
spin restricted LDA computations is calculated
with the VWN functional\cite{vwn}.

The Laplacian operator is discretized with a high order
finite difference representation\cite{hamming}.  
In the present work, a 12th order
form is used.  The high order expression yields a large gain in 
accuracy, which reduces the number of required grid points
significantly.  The same
Laplacian is used for solution of the Poisson and KS equations, 
resulting in a consistent level of accuracy throughout the solver.
Due to this consistent representation on all
levels, the only parameter controlling the accuracy
of the solution is the fine grid spacing $h$.  
The other quantities are diagonal in the coordinate representation.
The real space discretization scheme offers efficient calculation of electron densities and, as a result, the 
exchange-correlation potentials.
The nuclear charge density is the discretized form of the Dirac 
delta function.  The 
wavefunctions are vectors of length $N_g$, the total number of
grid points on a given level. 
Integration is performed by simple
trapezoidal summation on the three dimensional domains.    The boundary
conditions on the orbitals for the finite systems examined here
are set to zero, while the electrostatic potentials on the boundary
are obtained via multipole expansion of the total 
charge density to quadrupole order.  

The KS eigenvalue problem is nonlinear since one 
simultaneously solves 
for both the eigenvalues and the eigenfunctions.  Therefore,
we have employed the FAS eigenvalue technique of 
Brandt, {\it et al.}\cite{abeval} 
This method allows for solution of
nonlinear problems with efficiencies similar to linear
ones. Here we outline the FAS method of solution for
the eigenvalue and Poisson problems.
Both solvers are required
for the self consistent problem. 

The discrete equations can be represented
on the current finest grid as:

\begin{equation}
L^h U^h  = f^h.
\end{equation} 
 
\noindent
For the Poisson problem $L^h$ is the finite difference Laplacian
on the fine grid with spacing $h$,
$U^h$ is the electrostatic potential $v_{es}({\bf r})$, and
$f^h$ is $-4\pi\rho_{tot}({\bf r})$. Upper case for the potential
denotes the exact numerical solution. Below, lower case
implies the current approximation. In the 
eigenvalue equations $U^h$
becomes $U_i^h$, the eigenfunctions,
where the index $i$ indicates the orbital 
number.  The eigenvalue operator $L^h$ is the Hamiltonian minus
$\lambda_i$, where $\lambda_i$ is the eigenvalue for 
orbital $i$, and there is no source term $f^h$.  
The eigenvalue has the same value on all levels at convergence
and therefore is not labeled with a grid size. 

In the FAS method, the desired functions are represented 
on each grid level.  The discrete equations 
on the coarse level, however, are modified by the 
inclusion of a defect correction term which is required
to yield zero correction at convergence.
Suppose we have an approximate solution on the fine grid $h$, $u^h$,
which is obtained by relaxation sweeps on the grid:

\begin{equation}
L^h u^h = f^h.
\end{equation}

\noindent
We have used Gauss-Seidel or Successive Over-Relaxation
updates for these smoothing steps. 
The coarse grid problem with grid spacing
$H = 2h$ can then be constructed as: 

\begin{equation}
L^Hu^H = I^H_h f^h + \tau^H ,
\end{equation}

\noindent
where $I^H_h$ is the restriction operator which takes a 
local average of the fine grid function. Full weighting 
restriction is employed here, which weights the local 
points according to trapezoid rule integration. The
defect correction is: 

\begin{equation}
\tau^H = L^H I^H_h u^h - I^H_h L^h u^h .
\end{equation}

\noindent
The coarse grid equation is iterated to obtain 
an improved solution at that level.  In a two grid
method, the fine grid function is then corrected
via:
  
\begin{equation}
u^h \leftarrow u^h + I^h_H (u^H - I^H_h u^h) ,
\end{equation}

\noindent
where $I_H^h$ is the interpolation operator.  
These operations can be recursively extended to yet coarser
grids.  By inclusion of multiple levels, errors of all 
wavelengths are rapidly damped in the fine grid function.
The only change required on grids two or further from 
the finest grid is that an additional term $I_h^H \tau^h$ 
must be included in the defect correction to incorporate
information from the previous level. 

In solving the eigenvalue equations in KS theory, 
some additional components are required.  On all 
grids except the coarsest, the orbital equations
are updated just as outlined above. However, on 
the coarsest level, constraints must be imposed
to maintain wavefunction separation.  If one
had the exact solution on a fine grid $U_i^h$
and then restricted the orbitals to the next 
coarser level, the functions would no longer be
orthogonal. 
Hence, the constraints on the coarsest
level can be implemented by solving a linear
matrix equation:

\begin{equation}
\langle u_i^H, I^H_h u_j^h \rangle = \langle
I_h^H u_i^h , I_h^H u_j^h \rangle
\end{equation}

\noindent
These equations, since they are implemented only on the 
coarsest scale, require very little computational overhead,
and can be solved either via exact matrix inversion or by
iterative methods. The matrix size is of 
order $q \times q$, where
$q$ is the number of orbitals.   
The eigenvalue is also updated on the coarsest level 
by computing the Rayleigh quotient:

\begin{equation}
\lambda_i = \frac{<L^Hu_i^H-\tau_i^H,u_i^H>}{<u_i^H,u_i^H>}
\end{equation}

Finally, the convergence can be improved by 
including subspace orthogonalization via Ritz projection.
At the end of an MG correction cycle, a Gram-Schmidt 
orthogonalization is first performed on the finest level,
followed by diagonalization of the Hamiltonian in the basis
of the occupied orbitals:

\begin{equation}
\omega^T L \omega z_i -\lambda_i z_i = 0
\end{equation}

\noindent
where $\omega$ symbolizes the $q \times N_g$ matrix of
eigenfunctions on the grid, and the $z_i$ are the 
coefficients used to improve the occupied
subspace. This matrix problem 
is thus also 
of size $q \times q$. 

The FMG approach is initiated by first iterating the 
self consistent problem on the coarsest level used 
for the eigenvalue problem.  The initial approximation
for the eigenfunctions is a set of random numbers of magnitude
one.  
The coarsest level is chosen so 
that the eigenvalues have the same ordering as on 
the finest level. For the systems studied here, a coarse
grid of $8^3$ or $17^3$ is required, depending on the problem. 
The Poisson solver extends
the levels up to a $3^3$ grid where only the one central 
point is iterated.  In most of these calculations the 
finest grid employed is of size $65^3$, so the eigenvalue
solver comprises three or four levels, while a total of 
six levels is utilized for the Poisson solver on
the finest scale.  During the initial 
iterations on the coarest level, the eigenfunctions
are first relaxed and
orthogonalized via a Gram-Schmidt process, and then 
the potential is updated.  
Once an initial approximation is obtained on 
the coarse level, the 
solution is interpolated to the next finer level, where
the MG correction cycles are initiated. 

At the beginning 
of each MG V-cycle, the effective potential is updated
once upon entry to the new finest level with the 
FAS method.  Subsequently, the potential
is updated at the end of each eigenfunction correction cycle;  
since the changes to the potential are 
relatively smooth, the potential corrections are 
performed with a simple V-cycle using the previous potential
as input.  A 
schematic diagram of the MG solver is presented in Figure 1. 
The Poisson equation is typically solved with 
only a few relaxation
sweeps on the finest scale on either side of 
the V-cycle. During each eigenfunction
correction cycle, the eigenfunctions are relaxed 3 times
on each side of the V cycle. 
The solution of the 
entire electrostatic portion of the problem thus 
requires similar computational effort as for the update
of a single eigenfunction.  
This algorithm 
differs from those of Refs.\ \cite{bernholc,ancilotto}
by inclusion of updates of the eigenvalues and enforcement
of the constraints on the coarsest
level {\it only} without resorting to linearization of 
the equations.

We have explored two approaches for the eigenfunction
correction cycles.  In the first, each eigenfunction
is carried through the V cycle starting with
the lowest energy function, and the effective potential
is updated at the conclusion of the cycle.  This 
sequential method leads
to a rapidly convergent effective potential since the 
low lying states are quickly stabilized. In the second
approach, all eigenfunctions are corrected simultaneously,
and the effective potential is updated upon conclusion
of the V cycle.  For small systems the first approach
is preferred since the Poisson updates are quite 
inexpensive.  However, as we discuss below, the scaling
of the two approaches differs, and the second 
method is preferable for larger systems.  Its convergence
behavior is slightly less dramatic, requiring three or four 
self consistency updates as opposed to two for the 
first method.  The algorithm can readily be adapted 
for either approach depending on the problem. 

We have performed calculations on atoms, ions, and small 
molecules to test the convergence and accuracy of the method. 
The numerical results presented here were obtained with 
the second approach discussed above. 
The convergence behavior is illustrated in Fig.\ 2 
in calculations on
the 4 electron Be atom and the 14 electron CO 
molecule. Comparison is made to recent
MG and Car-Parrinello pseudopotential
calculations\cite{ancilotto} on the 8 valence 
electron C$_2$ molecule. 
The FAS-FMG approach leads to significant acceleration;
we emphasize that the present calculations include 
the nuclear singularity in the effective potential,
making it a more challenging scenario for 
convergence to the ground state.  
We computed total energies for
atoms and ions to obtain atomic
ionization potentials at the LDA-VWN
level (Table I).   The total 
energy obtained for the He atom is -2.8345 au {\it vs.}
the exact value of -2.8348 au. The computations
on the ions are performed with equal effort 
as for the neutral species, and the IP results
are of satisfactory accuracy\cite{tong}. 
Finally, 
we present the eigenvalues of the CO molecule
computed at the $X\alpha$ level for comparison
with previous numerical 
work\cite{velde} (Table I).  
The computation
was performed on a large 129$^3$ domain since the
dipole moment of CO is a sensitive function of 
the overall domain size. We obtained
a dipole value of 0.25 D 
C$^-$O$^+$ compared to the previous result
of 0.24 D in fully converged numerical 
calculations\cite{laaksonen}, and
0.10 D in finite difference pseudopotential 
calculations\cite{chel}. 
As discussed in Ref.\ \cite{wwt}, the most severe
errors in real space all electron calculations occur 
in the regions around the nuclei. The limitations
on the accuracy of our 
uniform domain results is a consequence 
of these errors.
Two improvements are
suggested: inclusion of pseudopotential 
techniques to remove the core and/or local grid 
refinements in the neighborhood
of the nucleus\cite{gygi,tbce}.  
We have generalized a second order multigrid 
local mesh refinement method\cite{bb} 
to arbitrary order and 
will include these refinements in the KS solver
in future work\cite{tbce}.  Since the 
refinements are truly local, they require only modest 
computational overhead. 

To conclude, we discuss the scaling properties of the 
various steps in the FAS-FMG algorithm for the case
where the orbitals span the whole domain. With $q$
orbitals and $N_g$ fine grid points in the domain, the
scalings of the important components of the algorithm are
as follows: relaxation of orbitals ($qN_g$), 
relaxation of potential ($N_g$), Gram-Schmidt 
process ($q^2 N_g$), Ritz projection ($q^2 N_g$
to construct the matrix and $q^3$ to solve), computation
of the eigenvalues on the coarse grid ($qN_g^H$), and
solution of the constraint equations on the coarse grid
($q^2N_g^H$ to construct the
constraint matrix and $q^3$ to solve). 
The most costly portion of the algorithm 
for the small systems examined here is the relaxation
step for the orbitals. For large systems where 
orbital localization is possible, each of the steps
becomes linear scaling with the number of electrons.
The two algorithms for implementing self consistency 
discussed above differ in that the first approach, while 
exhibiting stronger convergence to the ground state, 
leads to a $qN_g$ scaling which persists even with
localized orbitals.  

We have presented a new method for solution of
the KS equations in real space which utilizes 
nonlinear multigrid techniques to solve the 
self consistent equations.  The purpose has been 
to illustrate the rapid convergence behavior
of the FAS-FMG algorithm in relation to other 
electronic structure methods.  
The efficiency is a result of the 
preconditioning on coarser levels
and nonlinear treatment during the MG correction cycles.
The combination of real space discretization with
FAS-FMG multiscale techniques of solution leads to 
order of magnitude improvement in convergence
behavior relative to other MG methods and the 
Car-Parrinello approach, and it should 
thus prove a useful
numerical stategy for solving large scale electronic
structure problems. 
  

We would like to thank Prof.\ Achi Brandt for many helpful discussions concerning this work. The research was supported by NSF grant
CHE-9632309.


\begin{table}  
\begin{tabular}{lrrr}
&h&Ref. 16&FAS-FMG\\
\tableline
He$\rightarrow$ He$^+$& 0.15 & 0.969 & 0.974 \\
Li$\rightarrow$ Li$^+$& 0.35 & 0.198 & 0.193 \\
O$\rightarrow$ O$^{2+}$ & 0.12 & 1.843 & 1.821 \\
\end{tabular}
\begin{tabular}{lrr}
Orbital& Ref. 17&FAS-FMG \\
\tableline
1s(O) & -18.745& -18.832\\
1s(C) & -9.912 & -9.940\\
$\sigma$(2s) & -1.045& -1.060\\
$\sigma^*$(2s) & -0.489& -0.494\\
$\pi$(2p) & -0.413& -0.407\\
$\sigma$(2p) & -0.304& -0.302\\
\end{tabular}
\label{table1}
\caption{Calculations on Atomic Ionization Potentials 
and Eigenvalues
for the CO molecule. The LDA calculations of Ref. 16 used
a form for the correlation energy which interpolated
between the Wigner form for low densities and the 
Gell-Mann/Brueckner form at high densities.
For the CO calculation, the grid 
spacing was $h = 0.1335 au$ and the bond length was taken as 1.13\AA.
All energies are in hartrees.}
\end{table}

\begin{figure}
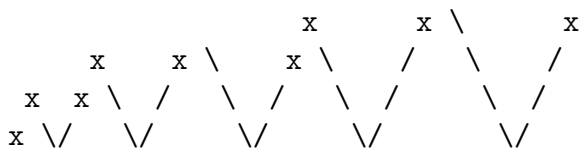

\begin{center}     
\begin{verbatim}

              
                  x      x \      x
     x    x \    x \    /   \    /  
 x  x \  /   \  /   \  /     \  /
x \/   \/     \/     \/       \/
       
\end{verbatim}
\end{center}
\caption{The FMG scheme. The fine grid 
solution is initiated with the interpolated functions from the 
coarse grid approximation(bottom of figure). 
At the end of each V cycle the potential is updated (indicated
by x).}
\end{figure}

\begin{figure}
\begin{center}
\setlength{\unitlength}{0.240900pt}
\ifx\plotpoint\undefined\newsavebox{\plotpoint}\fi
\sbox{\plotpoint}{\rule[-0.200pt]{0.400pt}{0.400pt}}%
\begin{picture}(1200,1028)(0,0)
\font\gnuplot=cmr10 at 10pt
\gnuplot
\sbox{\plotpoint}{\rule[-0.200pt]{0.400pt}{0.400pt}}%
\put(121.0,123.0){\rule[-0.200pt]{4.818pt}{0.400pt}}
\put(101,123){\makebox(0,0)[r]{-5}}
\put(1180.0,123.0){\rule[-0.200pt]{4.818pt}{0.400pt}}
\put(121.0,296.0){\rule[-0.200pt]{4.818pt}{0.400pt}}
\put(101,296){\makebox(0,0)[r]{-4}}
\put(1180.0,296.0){\rule[-0.200pt]{4.818pt}{0.400pt}}
\put(121.0,469.0){\rule[-0.200pt]{4.818pt}{0.400pt}}
\put(101,469){\makebox(0,0)[r]{-3}}
\put(1180.0,469.0){\rule[-0.200pt]{4.818pt}{0.400pt}}
\put(121.0,642.0){\rule[-0.200pt]{4.818pt}{0.400pt}}
\put(101,642){\makebox(0,0)[r]{-2}}
\put(1180.0,642.0){\rule[-0.200pt]{4.818pt}{0.400pt}}
\put(121.0,815.0){\rule[-0.200pt]{4.818pt}{0.400pt}}
\put(101,815){\makebox(0,0)[r]{-1}}
\put(1180.0,815.0){\rule[-0.200pt]{4.818pt}{0.400pt}}
\put(121.0,988.0){\rule[-0.200pt]{4.818pt}{0.400pt}}
\put(101,988){\makebox(0,0)[r]{0}}
\put(1180.0,988.0){\rule[-0.200pt]{4.818pt}{0.400pt}}
\put(178.0,123.0){\rule[-0.200pt]{0.400pt}{4.818pt}}
\put(178,82){\makebox(0,0){2}}
\put(178.0,968.0){\rule[-0.200pt]{0.400pt}{4.818pt}}
\put(291.0,123.0){\rule[-0.200pt]{0.400pt}{4.818pt}}
\put(291,82){\makebox(0,0){4}}
\put(291.0,968.0){\rule[-0.200pt]{0.400pt}{4.818pt}}
\put(405.0,123.0){\rule[-0.200pt]{0.400pt}{4.818pt}}
\put(405,82){\makebox(0,0){6}}
\put(405.0,968.0){\rule[-0.200pt]{0.400pt}{4.818pt}}
\put(519.0,123.0){\rule[-0.200pt]{0.400pt}{4.818pt}}
\put(519,82){\makebox(0,0){8}}
\put(519.0,968.0){\rule[-0.200pt]{0.400pt}{4.818pt}}
\put(632.0,123.0){\rule[-0.200pt]{0.400pt}{4.818pt}}
\put(632,82){\makebox(0,0){10}}
\put(632.0,968.0){\rule[-0.200pt]{0.400pt}{4.818pt}}
\put(746.0,123.0){\rule[-0.200pt]{0.400pt}{4.818pt}}
\put(746,82){\makebox(0,0){12}}
\put(746.0,968.0){\rule[-0.200pt]{0.400pt}{4.818pt}}
\put(859.0,123.0){\rule[-0.200pt]{0.400pt}{4.818pt}}
\put(859,82){\makebox(0,0){14}}
\put(859.0,968.0){\rule[-0.200pt]{0.400pt}{4.818pt}}
\put(973.0,123.0){\rule[-0.200pt]{0.400pt}{4.818pt}}
\put(973,82){\makebox(0,0){16}}
\put(973.0,968.0){\rule[-0.200pt]{0.400pt}{4.818pt}}
\put(1086.0,123.0){\rule[-0.200pt]{0.400pt}{4.818pt}}
\put(1086,82){\makebox(0,0){18}}
\put(1086.0,968.0){\rule[-0.200pt]{0.400pt}{4.818pt}}
\put(1200.0,123.0){\rule[-0.200pt]{0.400pt}{4.818pt}}
\put(1200,82){\makebox(0,0){20}}
\put(1200.0,968.0){\rule[-0.200pt]{0.400pt}{4.818pt}}
\put(121.0,123.0){\rule[-0.200pt]{259.931pt}{0.400pt}}
\put(1200.0,123.0){\rule[-0.200pt]{0.400pt}{208.378pt}}
\put(121.0,988.0){\rule[-0.200pt]{259.931pt}{0.400pt}}
\put(0,555){\makebox(0,0){log(Ediff)}}
\put(660,41){\makebox(0,0){number of SCF steps}}
\put(121.0,123.0){\rule[-0.200pt]{0.400pt}{208.378pt}}
\put(235,988){\usebox{\plotpoint}}
\multiput(235.00,986.92)(1.095,-0.500){671}{\rule{0.976pt}{0.120pt}}
\multiput(235.00,987.17)(735.974,-337.000){2}{\rule{0.488pt}{0.400pt}}
\put(121,891){\usebox{\plotpoint}}
\multiput(121,891)(16.073,-13.132){59}{\usebox{\plotpoint}}
\put(1061,123){\usebox{\plotpoint}}
\sbox{\plotpoint}{\rule[-0.400pt]{0.800pt}{0.800pt}}%
\put(121,653){\usebox{\plotpoint}}
\multiput(122.41,640.05)(0.502,-1.840){107}{\rule{0.121pt}{3.119pt}}
\multiput(119.34,646.53)(57.000,-201.526){2}{\rule{0.800pt}{1.560pt}}
\multiput(179.41,437.35)(0.502,-1.032){107}{\rule{0.121pt}{1.842pt}}
\multiput(176.34,441.18)(57.000,-113.177){2}{\rule{0.800pt}{0.921pt}}
\multiput(236.41,321.30)(0.502,-0.887){105}{\rule{0.121pt}{1.614pt}}
\multiput(233.34,324.65)(56.000,-95.649){2}{\rule{0.800pt}{0.807pt}}
\multiput(292.41,223.04)(0.502,-0.774){107}{\rule{0.121pt}{1.435pt}}
\multiput(289.34,226.02)(57.000,-85.021){2}{\rule{0.800pt}{0.718pt}}
\sbox{\plotpoint}{\rule[-0.500pt]{1.000pt}{1.000pt}}%
\put(121,485){\usebox{\plotpoint}}
\multiput(121,485)(3.272,-20.496){18}{\usebox{\plotpoint}}
\put(178,128){\usebox{\plotpoint}}
\end{picture}
\end{center}
\caption{Convergence behavior. 
The top curve is the Car-Parrinello result
of Ref. 8. The second curve is the MG 
result of Ref. 8. The next is our result
for the CO molecule with the FAS-FMG solver.
The bottom curve is the FAS-FMG result 
for the Be atom.}
\end{figure}
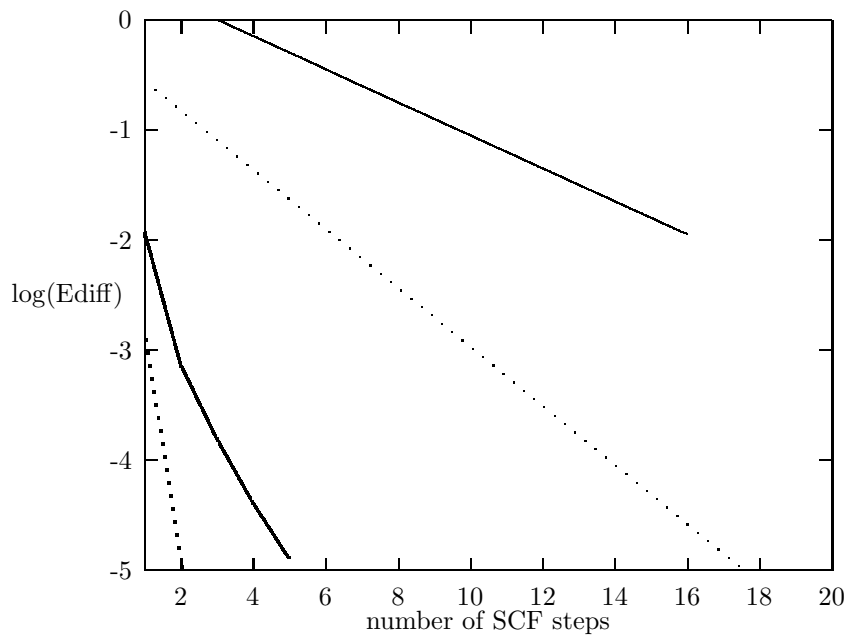

\end{document}